# Superconducting transitions of intrinsic arrays of weakly coupled 1D superconducting chains: The case of the extreme quasi-1D superconductor $Tl_2Mo_6Se_6$

B. Bergk[1][*], A.P. Petrović[2], Z. Wang[1], Y. Wang[1], D. Salloum[3], P. Gougeon[3], M. Potel[3], and R. Lortz[1][♠]

[1]*Department of Physics and William Mong Institute of Nano Science and Technology, Hong Kong University of Science & Technology, Clear Water Bay, Kowloon, Hong Kong*
[2]*Phase Control in Smart Materials Laboratory, Division of Physics and Applied Physics, School of Physical and Mathematical Sciences, Nanyang Technological University, 21 Nanyang Link, Singapore 637371*
[3]*Sciences Chimiques, CSM UMR CNRS 6226, Université de Rennes 1, Avenue du General Leclerc, 35042 Rennes Cedex, France*

**Abstract.** $Tl_2Mo_6Se_6$ represents a model system for quasi-one-dimensional superconductors. We investigate its superconducting transition in detail by means of electrical transport experiments on high-quality single crystalline samples with onset $T_c$ = 6.8 K. Our measurements indicate a highly complex superconducting transition which occurs in different stages, with a characteristic bump in the resistivity and distinct plateau structures in the supercurrent gap imaged by *V-I* curves. We interpret these features as fingerprints of the gradual establishment of global phase coherence in an array of weakly coupled parallel one-dimensional (1D) superconducting bundles. In this way, we demonstrate that superconducting $Tl_2Mo_6Se_6$ behaves like an intrinsic array of Proximity or Josephson junctions, undergoing a complex superconducting phase-ordering transition at 4.5 K which shows many similarities to the Berezinskii-Kosterlitz-Thouless transition.

## 1. Introduction

Following the recent emergence of novel nanomaterials, understanding the impact of reduced dimensionality on fundamental physical properties is of high technological interest. The presence of thermal and quantum fluctuations will prevent a strictly 1D system with short-range interactions from developing long range order at finite temperatures [1]. This limitation may however be overcome if such 1D objects are arranged in an array of parallel 1D chains. If the interchain coupling in such an array becomes sufficiently strong, the array may undergo a phase transition into a state with long-range order at a finite temperature [2]. One prominent example is offered by the family of quasi-1D superconductors [3,4]. A quasi-1D superconductor is generally defined as a system in which the thickness of either an individual nanowire or a metallic atomic chain within a crystallographic structure is smaller than the Ginzburg-Landau coherence length. Superconductivity in a strictly 1D system is not possible [5] and therefore, in order to develop a true bulk superconducting state, the quasi-1D building blocks are required to have a certain minimum thickness or transversal coupling to neighbouring parallel chains within the array. Early examples of crystalline compounds exhibiting intrinsically quasi-1D superconductivity were found in the family of highly

---

[*] present address: Hochfeld-Magnetlabor Dresden, Forschungszentrum Dresden-Rossendorf, 01314 Dresden, Germany

[♠] corresponding author: lortz@ust.hk

anisotropic organic Bechgaard salts such as $(TMTSF)_2PF_6$ [6]. With the emergence of nanofabrication techniques, quasi-1D superconductors can be tailored in the form of individual metallic nanowires [7] or arrays [8], as well as in matrices containing parallel ultra-thin carbon nanotubes [9-11].

$Tl_2Mo_6Se_6$ is a further example of a crystalline structure in which quasi-1D superconductivity develops intrinsically. It is formed from $Mo_6Se_6$ "chains" of theoretically infinite length, weakly coupled by $Tl^+$ ions in the channels between the chains (Figure 1) [12]. Related to the Chevrel phases, it remains one of the most strongly anisotropic quasi-one-dimensional superconductors currently known, with $T_c$ varying between 3 K and 6.5 K [13]. To its detriment, $Tl_2Mo_6Se_6$ was discovered shortly before the appearance of the high-$T_c$ cuprate superconductors; since then it has been largely shunned by the scientific community in favour of more glamorous compounds with higher critical temperatures. However, as we demonstrate in our paper, this seems to have been a significant oversight.

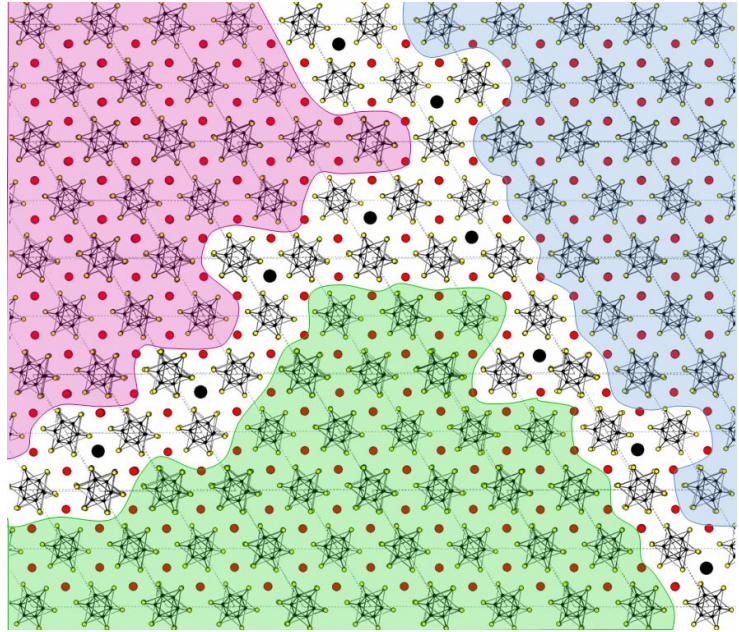

**Figure 1.** Structure of a typical $Tl_{2-x}Mo_6Se_6$ crystal in the *ab* plane lateral to the 1D $Tl_2Mo_6Se_6$ chains. The $Mo_6Se_6$ chains which intercept this plane are represented by the clusters formed of small black and yellow dots. Red dots represent the Tl atoms. The different shaded areas represent the Tl-rich islands (or bundles) that become superconducting separated by (unshaded) metallic Tl deficient regions.

In low-dimensional superconducting materials (i.e. those with dimensionality less than 3), thermal and quantum phase fluctuations of the order parameter are enhanced. This should in principle reduce the true transition temperature (which corresponds to the onset of zero resistivity) to zero [5]. Nevertheless, a two-dimensional (2D) superconductor has the possibility to overcome this limitation by undergoing the famous Berezinskii-Kosterlitz-Thouless (BKT) transition at finite temperatures: this represents a phase-ordering transition within the 2D plane between a phase-incoherent high-temperature and a phase-coherent low-temperature superconducting state.

Tl$_2$Mo$_6$Se$_6$ is however neither a 2D superconductor nor a perfectly 1D superconductor (which would indeed exhibit $T_c = 0$). Molybdenum selenide chain compounds do not generally exhibit perfect stoichiometry: in Tl$_2$Mo$_6$Se$_6$ the Tl content per formula unit is known to vary from 1.9 to 2 (and hence Tl$_{2-x}$Mo$_6$Se$_6$ could be argued to be a more appropriate formula). This implies that within a single Tl$^+$ channel, one could expect a minimum average vacancy spacing of roughly 5nm (assuming a random defect distribution). It follows that the presence of any Tl vacancies between the chains should transform the material into a heterogeneous network of Tl$_2$Mo$_6$Se$_6$ islands separated by narrow Mo$_6$Se$_6$ regions with quasi-1D characteristics due to the large anisotropy (as represented in Figure 1). Since Tl$_2$Mo$_6$Se$_6$ becomes superconducting at low temperature while Mo$_6$Se$_6$ remains metallic, a typical crystal will be composed of superconducting bundles of Tl$_2$Mo$_6$Se$_6$ chains, several tens of nanometers in diameter, separated by narrow Tl-deficient metallic Mo$_6$Se$_6$ zones. (Note that it is unclear whether any microscopic phase separation occurs in Tl$_{2-x}$Mo$_6$Se$_6$: this would both increase the size and decrease the number of the superconducting and metallic regions).

Any nanoscale intergrowth of metallic and superconducting regions generates intrinsic arrays of Proximity junctions (PJ) or Josephson junctions (JJ) which will strongly influence the superconducting properties of such materials [14,15]. In the last three decades the behavior of such arrays has been studied in detail [15-21] as ideal model systems for real materials, mainly to gain a deeper insight into the nature of the phase transition from the normal into the superconducting state. In these materials, the bulk superconducting properties are strongly influenced by fluctuations in the phase of the order parameter. Intrinsic Josephson junctions (JJ) are known to play an important role in the physics of layered superconductors such as high-$T_c$ cuprates [22].

Due to the logarithmic divergence in the phase fluctuations at long wavelengths, the transition into the superconducting state of PJ and JJ arrays belongs to the BKT universality class. In general, a BKT transition is a thermally driven vortex-antivortex unbinding transition of infinite order within the theoretical framework of the 2D-XY model. Below the transition temperature $T_{BKT}$, the vortices and antivortices form bound pairs and trigger spontaneous macroscopic phase coherence. Above $T_{BKT}$ but below the onset temperature of superconducting correlations, the vortex-antivortex pairs dissociate and local phase slips related to unbound vortices destroy global phase coherence [23-25]. This regime is characterized by a non-zero electrical resistance, although local phase coherence exists over short length and timescales. For electrical transport experiments, theory predicts two characteristic signatures of such a transition. First, coming from temperatures above the phase transition the resistance $R$ follows the thermodynamic scaling relation $R = A \exp(b/(T_{BKT}-1)^{-1/2})$ ($A$, and $b$ are numerical constants) [16,26,27] which reflects the binding of the vortex pairs. The second property can be observed in the $V$-$I$-characteristics. The power-law exponent $\alpha(T)$, where $V \propto I^{\alpha(T)}$ provides the principal means of identification for a BKT-like transition, since it is predicted to jump from 1 at high temperatures above the transition to larger values with decreasing temperature. The resulting discontinuity is known as the Nelson-Kosterlitz jump [28] and at $T_{BKT}$ $\alpha(T)$ is expected to be precisely 3.

As a model system for low-dimensional superconductors, quasi-one-dimensional Tl$_2$Mo$_6$Se$_6$ allows us to study the importance of intrinsic JJ and BKT behaviour in a single crystalline material rather than a synthetic array. Although BKT behaviour is generally expected to occur in 2D superconductors, arrays of parallel 1D superconducting bundles may undergo a similar phase transition of the same universality class: superconducting condensates are initially formed in the individual bundles, although the longitudinal phase coherence is disturbed by strong longitudinal phase fluctuations. At this stage, there is no lateral phase coherence between bundles. As the temperature falls, the transverse coupling of the bundles via the Josephson effect will trigger phase coherence within the 2D lateral plane which instantly stabilizes the longitudinal phase and restores a 3D bulk superconducting state. The BKT vortices are formed by excitations or slips in the phase of the order parameter between neighboring bundles in this plane.

Motivated by recent work reporting strong evidence that arrays of superconducting parallel 4 Ångstrom carbon nanotubes grown in the linear channels of $AlPO_4$-5 zeolite undergo precisely this type of phase transition [9], we have identified $Tl_2Mo_6Se_6$ as a potential candidate to undergo a similar superconducting transition. In this paper, we discuss the electrical transport properties of $Tl_2Mo_6Se_6$ within the framework of a BKT transition. Remarkably, both CNT [10] and $Tl_2Mo_6Se_6$ [29] show a very similar superconducting specific-heat anomaly, with the characteristic broad hump of a BKT-type transition [26]. We also demonstrate that the electrical transport characteristics for both these superconductors are very similar and that the phase ordering part of the superconducting transition can be interpreted by analogy with the BKT transition in a 2D-XY spin system.

## 2. Experimental

Needle-like crystals of dimensions approximately 4 mm × 300 μm × 100 μm and mass 800 μg were synthesised in a sealed molybdenum crucible at 1700 °C. In Ref. [29] the properties of the samples have been investigated and discussed in detail. All measurements were carried out in a $^4$He cryostat at temperatures down to 2 K, with the electrical gold wire contacts attached to the samples using silver paste. In this manuscript we present measurements on two different crystals (Samples 1 and 2) which behave similarly. The majority of the data were taken in helium exchange gas, although we repeated several I-V curve measurements directly in liquid helium in order to rule out artifacts related to local heating effects at the sample contacts due to high currents. Both methods produced the same results.

## 3. Results

Figure 2 shows the temperature dependence of the longitudinal resistance of Sample 1 measured in different magnetic fields aligned perpendicular to the *c* (one-dimensional) axis of the sample. The zero-field resistance drops at 6.8 K and decreases gradually towards lower temperature, with a bump structure around 5.5 K. In non-zero fields the transition is shifted to lower temperatures but the characteristic shape of the cool-down curve is retained. Above roughly 2.5 T the superconductivity is suppressed completely. Similar behaviour has already been observed in other $Tl_2Mo_6Se_6$ samples and was attributed to a distribution of $T_c$ values due to stoichiometric variations, combined with a small misalignment of the applied field [29]. However, as we shall demonstrate the bump is a natural consequence of a BKT transition in this material.

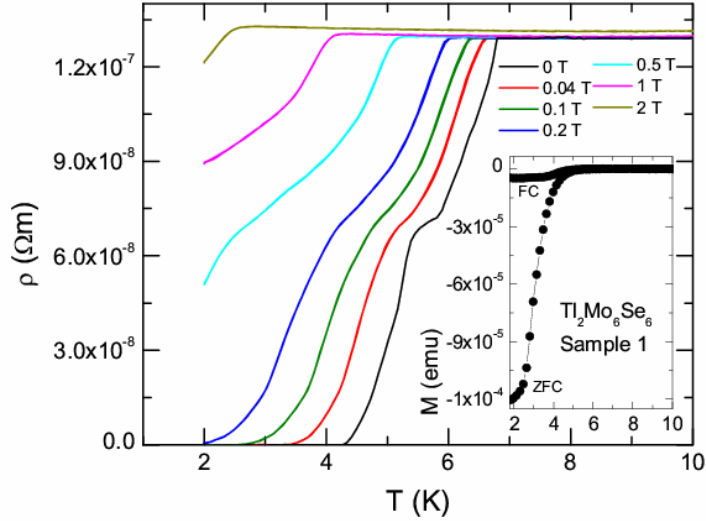

**Figure 2.** Temperature dependence of the electrical resistance of Sample 1 obtained at different magnetic fields perpendicular to the $c$ axis. Inset: Temperature dependence of the field (2 Gauss) and zero-field-cooled magnetization.

Upon comparing our results with transport measurements on PJ arrays, substantial similarities are immediately obvious, in particular the bump in the cool-down curve [16,17,30]. In the temperature region of the bump the proximity effect between the superconducting islands and non-superconducting metallic parts of the sample is strongly enhanced. The Josephson coupling between the superconducting islands only becomes important at temperatures below the bump [16,19] and complete phase coherence is only obtained when the resistance reaches zero. A similar model can immediately be assigned to $Tl_2Mo_6Se_6$ due to its one-dimensional crystal structure and non-stoichiometric nature. At the onset of the transition, small bundles of stoichiometric $Tl_2Mo_6Se_6$ become superconducting, while neighbouring Tl-deficient channels within the crystal remain metallic. As the temperature is reduced further and we reach the bump structure in $\rho(T)$, superconductivity is induced within these metallic $Mo_6Se_6$ regions via the proximity effect. Finally, global phase coherence is established and the resistance reaches zero. It should be noted that the strong anisotropy in the coherence length ($\xi_{//}$ = 94nm, $\xi_\perp$ = 7.5nm) enhances phase stiffness along the chain axis and hence accentuates the predisposition of the material to a BKT-style transition. Furthermore, the short lateral coherence length implies that Cooper pairs are able to reside within a single superconducting $Tl_2Mo_6Se_6$ bundle. This is supported by the absence of any Luttinger effects in the normal-state electrical transport, which would be expected in the extreme one-dimensional limit of decoupled individual chains [29].

Figure 3 and 4 show typical four-point resistance data of the two samples at various temperatures. A supercurrent gap is clearly visible in both samples, with the voltage (resistance) reaching zero at low currents for temperatures below about 4.5 K. Instead of a single kink as expected, the differential resistance d$V$ / d$I$ has a multiple peak structure when the gap is approached (Figure 3). Comparable behavior has been observed previously in JJ and PJ arrays: essentially, there are two possible explanations for these peaks. Firstly, they might result from a superposition of different junctions as seen for instance in $YBa_2Cu_3O_7$ JJ arrays [31]. Secondly, the peaks could originate from the so-called subharmonic gap structure which is a phenomenon in superconductor-metal-superconductor assemblies attributed to multiple Andreev reflections [32,33]. For currents below the energy gap in a PJ, the transport of quasiparticles between superconducting areas is limited. However, the quasiparticles gain energy via several Andreev reflections until they can overcome the barrier. Experimentally, these subharmonic gaps are observable as dip structures in the d$V$ / d$I$ curves at integer fractions of the gap value. This has already been observed in many SNS and Superconductor-Semiconductor-Superconductor junctions [34-39]. The analysis of the peak structure in our data is not as straightforward as for the artificial structures, since for our samples the precise assembly of the one-dimensional needles is not known. However, as the peaks in our $Tl_2Mo_6Se_6$ data seem to follow a similar temperature dependence, this provides a good explanation for our data.

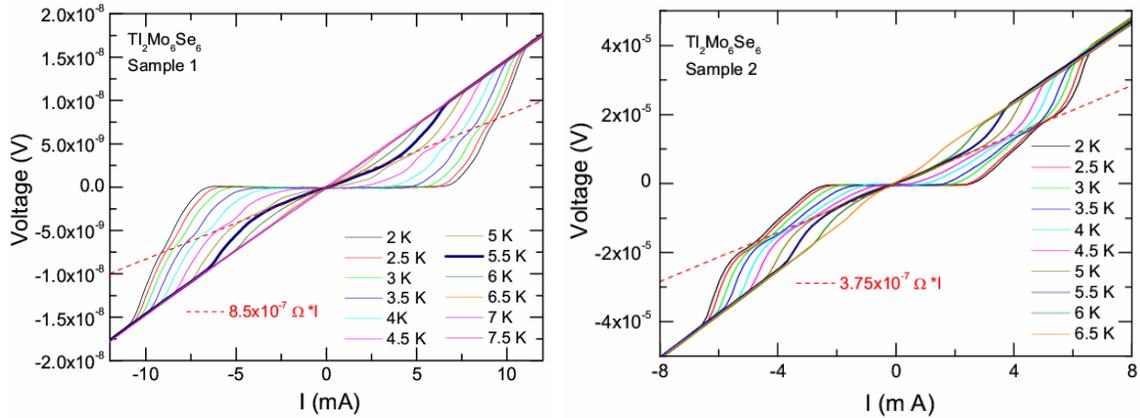

**Figure 3.** Voltage versus current for Sample 1 (left panel) and Sample 2 (right panel) measured for different temperatures. The dashed line represents the plateau resistance value in Figure 4.

Towards higher temperatures the supercurrent gap becomes smaller, until above 4.5 K the resistance no longer reaches zero. At 5.5 K a second plateau in the d$V$/d$I$ curves is attained in both samples (dashed lines in Fig. 4). In the *V-I* curves in Figure 3, a dashed line whose slope corresponds to the plateau resistance has been added: in contrast with higher-temperature data, curves acquired at T < 5.5 K cross this line twice at non-zero I, changing their slope at the intersection. This plateau temperature 5.5 K coincides with the bump in the resistive transition (Figure 2), suggesting that the plateau is representative of a pseudogap originating from local phase coherence in quasi-1D superconducting filaments. A true zero-resistance supercurrent gap may only be observed upon the establishment of global phase coherence, which occurs progressively below 5.5 K as the filaments are coupled laterally via the proximity effect.

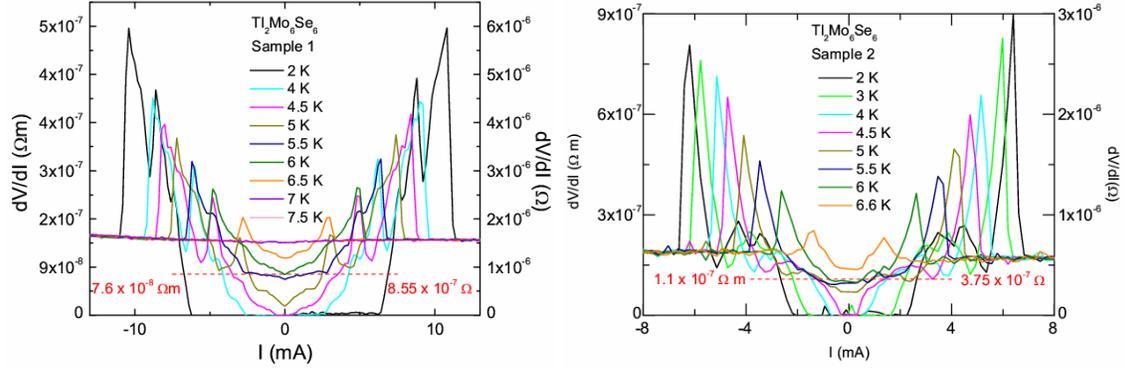

**Figure 4.** Differential resistance d$V$ / d$I$ as a function of the current $I$ at different temperatures for Sample 1 (left panel) and Sample 2 (right panel). The left scale is in units of the resistivity (Ωm) and the right scale in resistance (Ω). The data corresponds to the measurements shown in Fig. 3. The dashed line represents the second resistance plateau originating from the establishment of local phase coherence, prior to a BKT transition resulting in global phase coherence.

The shape of the observed gap transforms strongly with increasing temperature, from a more rounded shape at low temperature to a cusp-like structure at higher temperatures. Such behavior is a direct consequence of the temperature dependence of the *V-I*-curve power-law exponent $\alpha(T)$, where $V \propto I^{\alpha(T)}$. Unfortunately, many factors (including finite-size effects or free vortices [18,40]) may affect the *V-I* curves and hamper an accurate determination of $\alpha(T)$. This is illustrated in Figure 5, in which the *V-I* curves of Sample 1 are plotted on a double logarithmic scale. The straight lines represent power-law fits to the data. At high temperatures above 5 K the data appear to follow a power law; however in the temperature range between 4 and 5 K the slope changes strongly upon decreasing the temperature. Towards low temperature, the slope becomes steeper for small currents and only follows a power law within a small range, with a large value for the exponent $\alpha$. The multiple peak structure of the gap also impedes the curve analysis at low temperature, leading to a growth in the error bar of the $\alpha$-exponent as the temperature falls. The exponent $\alpha(T)$ for both samples is shown in Figure 5b. Above the onset of the resistive superconducting transition $\alpha$ is exactly 1, but below the onset temperature $\alpha$ starts to grow steadily towards low temperatures. Between 5 K and 3 K $\alpha$ rises more strongly with decreasing temperature, before finally tending towards saturation at the lowest temperatures measured. For Sample 1 we determine $T_{BKT}(\alpha=3) = 4.5 \pm 0.3$ K, while for Sample 2 $T_{BKT}(\alpha=3) = 4.2 \pm 0.3$ K. These temperatures correspond to the vanishing of the resistance (Figure 2) and confirm that the superconducting phase transition of $Tl_2Mo_6Se_6$ has a BKT-like phase ordering character. These slightly differing temperatures may be attributed to the arrangements of the fibers in each sample, which will certainly not be identical. In previous studies of PJ junction arrays it has been shown how this could influence the vortex binding transition [18].

The small discrepancy between the two samples is presumably due to different microscopic assemblies of $Mo_6Se_6$ chains along the exact current path probed during our transport experiments. Of course, in a single crystal of complex structure, it is difficult to achieve perfect conditions similar to those of an artificially produced junction array. The *V-I* curves depend critically on the assembly geometry and their behavior can hence vary wildly as a function of the junction array size [18]. As already pointed out, the local quality of the sample within the current path probed during the experiment is very important for observing pure BKT behavior. Therefore, the theoretically expected characteristics can easily be smeared out in the event of a superposition of slightly different array types.

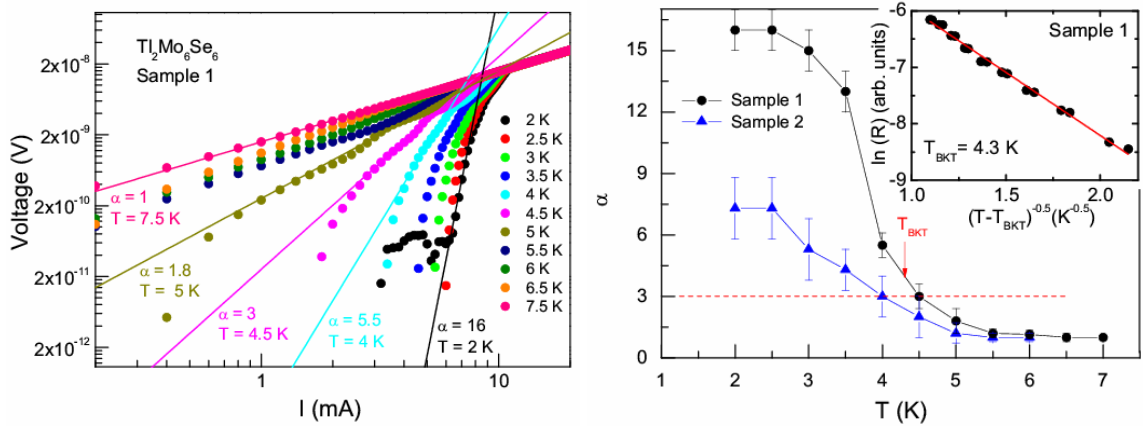

**Figure 5.** (a) *V-I* curve analysis in Sample 1. The straight lines represent the power-law fit $V \propto I^{\alpha(T)}$ to the data to determine $\alpha(T)$. (b) $\alpha(T)$ versus temperature $T$ for Samples 1 and 2. The inset illustrates the scaling of the resistance as a function of temperature for Sample 1.

Another method of identifying a BKT-like transition is given by the unique temperature scaling of the physical properties just above $T_{BKT}$ [26,41] which reflects the thermal activation of the unbound vortices and antivortices [16,30,42]. In this temperature range, the resistance should vary as $R = A \exp(b/(T_{BKT}-1)^{-1/2})$ where $A$ and $b$ are material constants. This is depicted in the inset of Figure 5b for sample 1. Using $T_{BKT} = 4.3$ K, the correct scaling behavior can clearly be verified.

The complex shape of the *V-I* curves, the scaling of the resistance and the jump in the exponent $\alpha$ strongly suggest that superconductivity appears in different stages in $Tl_2Mo_6Se_6$ upon lowering the temperature. At the upper onset of the resistive transition Cooper pairs are formed within individual $Tl_2Mo_6Se_6$ chains, but longitudinal fluctuations destroy macroscopic phase coherence. A transverse coupling of the chains via the proximity effect is required to stabilize the longitudinal phase and this coupling occurs at lower temperature in the form of a phase ordering transition within the plane perpendicular to the chains. A schematic diagram illustrating vortex formation via an array of phase-incoherent superconducting islands may be seen in Figure 6: such vortex excitations are bound with anti-vortices at $T_{BKT}$ to form a phase-coherent ground state.

This phase ordering transition shows many similarities to the BKT transition although $Tl_2Mo_6Se_6$ does not exhibit the standard 2D geometry in which the BKT transition is usually considered. Our experimental observations are however in excellent agreement with the various predictions of the BKT model. The superconducting transition displays similar characteristics to those observed in PJ arrays [16-18], thus demonstrating the existence of weak conducting links between the one-dimensional chains. In Table 1 we provide estimates of several relevant physical parameters based on this physical interpretation of our data, which are in excellent agreement with those of other quasi-1D superconductors [9].

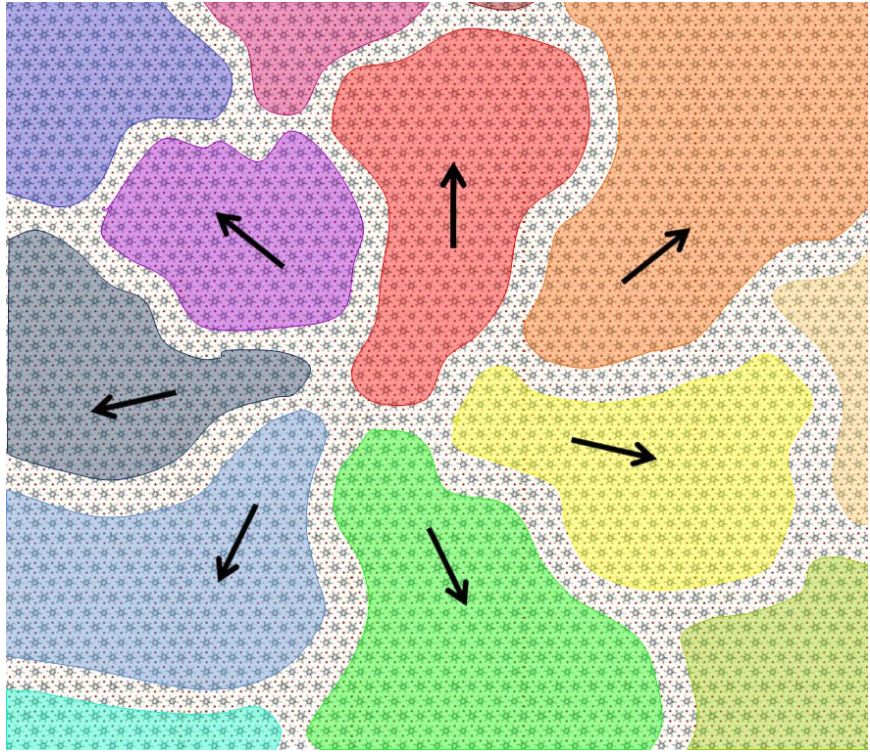

**Figure 6.** Vortex formation in a quasi-1D inhomogeneous superconductor, viewed along the *c* axis. Different shadings represent different phase values of the superconducting order parameter in distinct superconducting islands separated by (unshaded) metallic regions. An anti-vortex is formed in an identical manner, except that the phase rotates in the opposite direction (i.e. the arrows would point towards the centre of the diagram rather than outwards).

| | |
|---|---|
| *Josephson coupling energy J:* | From $\pi J / k_B T_{BKT} \approx 1.12$ [11] and $T_{BKT} = 4.5$ K: $\Rightarrow J = 0.14$ meV |
| *Critical Josephson current* $I_c = 2eJ/\hbar$ | $I_c = 0.0668$ μA |
| *Normal resistance $R_n$ per square in the transverse plane:* | From $\dfrac{T_{BKT}}{T_{c0}} = [1+0.173 R_n / (\hbar/e^2)]^{-1}$ [34] and $T_{BKT} = 4.5$ K, $T_{C0} = 6.8$ K: $\Rightarrow R_n = 8096$ Ω/□ |

**Table 1.** Estimates of several relevant physical parameters.

## 4. Conclusions

In summary, the superconducting transition in Tl$_2$Mo$_6$Se$_6$ displays very similar experimental characteristics to Berezinskii-Kosterlitz-Thouless behavior, with individually fluctuating 1D superconducting correlations preceding a transition to a coherent 3D bulk superconductor via a 2D phase-ordering transition. A broad hump in the specific heat [26,29], a bump-like anomaly within the resistive transition, multiple peaks and plateaus in the differential resistivity and a *V-I* curve critical exponent α = 3 have all been observed. $Tl_2Mo_6Se_6$ may therefore be considered to be a model system for low dimensional crystalline materials exhibiting intrinsic "weak links" between superconducting fibers or chains. Furthermore, the applicability of BKT physics to this compound suggests a new universality in the superconducting transitions of such materials.

**Acknowledgements**
We thank P. Sheng for fruitful discussions which stimulated this experimental work and U. Lampe for technical support. This work was supported by the Research Grants Council of Hong Kong Grants DAG08/09.SC01, SEG_HKUST03 and HKUST9/CRF/08.

**References**
[1] Landau L D and Lifschitz E M, *Statistical Physics* (Pergamon, New York, 1958).
[2] Scalapino D J, Imry Y, Pincus P, *Phys. Rev.* B **11**, 3042 (1975).
[3] Schulz H J, Bourbonnais C, *Phys. Rev. B* **27**, 5856 (1983).
[4] For a recent theoretical review see e.g. Arutyunova K Y, Golubevc D S, Zaikin A D, *Physics Reports* **464** , 1 (2008).
[5] Mermin N D, Wagner H, *Phys. Rev. Lett.* **17**; 1133 (1966), Hohenberg P C, *Physical Review* **158**, 383 (1967).
[6] For a review see: Bourbonnais C and Jérome D, *Interacting Electrons In Quasi-One-Dimensional Organic Superconductors,* Chapter of *The Physics of Organic Superconductors and Conductors*, Springer Series in Materials Science Vol. 110, edited by A. Lebed, (Springer, Heidelberg, 2008), p. 357-414.
[7] see e.g. Michottea S, Piraux L, Boyer F, Ladan F R, Maneval J P, *Appl. Phys. Lett.* **85**, 3175 (2004); Wang J, Ma X-C, Qi Y, Ji S-H, Fu Y-S, Lu L, Jin A-Z, *Appl. Phys. Lett.* **106**, 034301 (2004) and references therein.
[8] see e.g. Xu K and Heath J R, *Nano Letters* **8**, 136 (2008) and references therein.
[9] Wang Z, Shi W, Xie H, Zhang T, Wang N, Tang Z, Zhang X, Lortz R, and Sheng P, *Phys. Rev.* B **81**, 174530 (2010).
[10] Lortz R, Zang Q, Shi W, Ye J, Qiu C, Wang Z, He H, Sheng P, Qian T, Tang Z K, Wang N, Zhang X X, Wang J, Chan C T, *PNAS* **106**, 7299 (2009).
[11] Ieong C, Wang Z, Shi W, Wang Y, Wang N, Tang Z, Sheng P, Lortz R, *Phys. Rev.* B **83**, 184512 (2011).
[12] Potel M, Chevrel R, Sergent M, Armici J C, Decroux M and Fisher Ø, *J. Solid State Chem.* **35**, 286 (1980).
[13] Armici J C, Decroux M, Fisher Ø, Potel M, Chevrel R and Sergent M, *Solid State Commun.* **33**, 607 (1979).
[14] Kleiner R and Mueller P, *Phys. Rev.* B **49**, 1327 (1997).
[15] Sugano R, Onogi T, and Murayama Y, *Phys. Rev.* B **48**, 13784 (1993).
[16] Abraham D W, Lobb C J, Tinkham M, and Klapwijk T M, *Phys. Rev.* B **26,** 5268 (1982).
[17] Resnick D J, Garland J C, Boyd J T, Shoemaker S, and Newrock R S, *Phys. Rev. Lett.* **47**, 1542 (1981).
[18] Herbert S T, Jun Y, Newrock R S, Lobb C J, Ravindran K, Shin H-K, Mast D B and Elhamri S, *Phys. Rev.* B **57**, 1154 (1998).
[19] Martinoli P and Leeman C, *J. Low Temp. Phys.* **118**, 699 (2000).
[20] Capriotti L, Cuccoli A, Fubini A, Tognetti V, Vaia R, *Fundamental Problems of Mesoscopic Physics, NATO Science Series*, 2004, Volume 154, II, 203-216.
[21] Shaw T J, Ferrari M J, Sohn L L, Lee D-H, Tinkham M, and Clarke J, *Phys. Rev. Lett.* **76**, 2551 (1981).
[22] For a review see e.g. Buchanan M, *Nature* (London) **409**, 8 (2001).
[23] Kosterlitz J M and Thouless D J, *J. Phys.* C **6**, 1181 (1973).
[24] Kosterlitz J M, *J. Phys.* C **7**, 1046 (1974).
[25] Berezinskii V L, *Zh. Eksp. Teor. Fiz.* **59**, 207 (1970), *Sov. Phys. JETP* **32**, 493 (1971).
[26] Ding H-Q, *Phys. Rev.* B **45**, 230 (1996).

[27] Medvedyeva K, Beom Jun Kim, and Minnhagen P, *Phys. Rev.* B **62**, 14531 (2000).
[28] Nelson D R and Kostelitz J M, *Phys. Rev. Lett.* **39**, 1201 (1977).
[29] Petrović A P, Lortz R, Santi G, Decroux M, Monnard H, Boeri L, Andersen O K, Kortus J, Salloum D, Gougeon P, Potel M, Fischer Ø, *Phys. Rev.* B **82**, 235128 (2010) ;
Petrović A P, Fasano Y, Lortz R, Decroux M, Potel M, Chevrel R, Fisher Ø, *Physica* C **460-462**, 702-703 (2007).
[30] Kimhi D, Leyvraz F, and Ariosa D, *Phys. Rev.* B **29**, 1487 (1984).
[31] Burckhardt H, Rauther A, Schilling M, *Physica* C **326–327**, 93–98 (1999).
[32] Andreev A F, *Zh. Eksp. Teor. Fiz.* **46**, 1823 (1964).
[33] Klapwijk T M, Blonder G E, Tinkham M, *Physica* **109 & 110B**, 1657 (1982).
[34] Van Huffelen W M, Klapwijk T M, Heslinga D R, de Boer M J, and van der Post N, *Phys. Rev.* B **47**, 5170 (1993).
[35] Takayanagi H, *Physica* B **227**, 224-228 (1996).
[36] Schaepers, T H, Neurohr, K, Malindretos J, Kaluza A, Picard J-M, Lüth H, *Superlattices and Microstructures*, **25** (1999).
[37] Kutchinsky J, Taboryski R, Clausen T, Sørensen C B, Kristensen A, Lindelof P E, Bindslev Hansen J, Schelde Jacobsen C, and Skov J L, *Phys. Rev. Lett.* **78**, 931 (1997).
[38] Baturina T I, Islamov D R, Kvon Z D, *JETP Letters* **75**, 397 (2002).
[39] Cuevas J C, Hammer J, Kopu J, Viljas J K, and Eschrig M, *Phys. Rev.* B **73**, 184505 (2006).
[40] Chen Q-H, Tang L H, and Tong P, *Phys. Rev. Lett.* **87**, 067001 (2001).
[41] Halperin B I, and Nelson D R, *J. Low Temp. Phys.* **36,** 599 (1979).
[42] Beasley M R, Mooij J E, and Orlando T P, *Phys. Rev. Lett.* **42**, 1165 (1979).